\documentstyle[12pt]{article}
\normalbaselines
\def\beq{\begin{equation}}
\def\eeq{\end{equation}}

\begin{document}

\bigskip\bigskip

INFN-NA-IV-93/31~~~~~~~~~~~~~~~~~~~~~~~~~~~~~~~~~~~~~~~~~~~~~~DSF-T-93/31

%\vspace{4cm}

%\vbox{\vspace{38mm}}
%\begin{center}
\title{
{\normalsize ~~~~INFN-NA-IV-93/31 \hfill~~ DSF-T-93/31~~~~}
\vskip1.5cm
Thermal noise and oscillations of photon distribution
for squeezed and correlated light\\[2mm]}
%\end{center}
%\begin{center}
\author{V.V.Dodonov, O.V.Man'ko \\
{\it Lebedev Physics Institute, Moscow, Russia} \\
\and
V.I.Man'ko\thanks{ \it On leave of absence from
Lebedev Physics Institute, Moscow, Russia}~ and L.Rosa \\
{\it Dipartimento di Scienze Fisiche Universita di Napoli "Federico II" }\\
{\it and I.N.F.N.,Sez.di Napoli}}
%\end{center}

\maketitle

\begin{abstract}

Abstract:The oscillations of photon distribution function for squeezed
and correlated light are shown to decrease when the temperature
increases.The influence of the squeezing parameter and photon
quadrature correlation coefficient on the photon distribution
oscillations at nonzero temperatures is studied. The connection
of deformation of Planck distribution formula with oscillations of
photon distribution for squeezed and correlated light is discussed.
\end{abstract}

\newpage

\section{Introduction}

The density matrix of one-mode Gaussian nonclassical light may be
related to the problem of minimization of the Heisenberg uncertainty
relation \cite{Heis} and the Shcr\"odinger uncertainty relation
\cite{Schr},\cite{Robe}.Thus for the squeezed light \cite{Sto},
\cite{Hol} the product of quadrature dispersions is equal to the
lowest limit of the Heisenberg inequality.In Ref. \cite{Kur} the
correlated state of light has been found to minimize the
Schr\"odinger inequality.The squeezed nonclassical light differs
from the coherent state light \cite{Gla} in particular by the
properties of photon distribution function.Thus for coherent light
it is usual Poisson distribution which is smooth function with one
maximum near the mean of photon number.

It is shown in Refs.\cite{Sch1},\cite{Sch2} that the photon
distribution function of the squeezed light for some values of
squeezing parameters has strongly oscillating behaviour.In Ref.
\cite{Kli} the oscillatory character of photon distribution
was demonstrated for the correlated light.Squeezed and correlated
states as well as coherent states are the partial cases of the
Gaussian light whose density matrix in Wigner representation
is described by generic Gaussian function determined by five
real parameters which are two photon quadrature means $<p>$,
$<q>$ and three elements of quadrature dispersion matrix
$\sigma _{pp}$,$\sigma _{qq}$,$\sigma _{pq}$.The distribution
function for such generic Gaussian density matrix has been
expressed in terms of Hermite polynomials of two variables
in \cite{Sem},\cite{Vou},\cite{Dod1}.

In this work we address the question how the thermal noise
influences the photon distribution function
oscillations for squeezed and correlated light.
The aim of this work is to show explicitly that photon
distribution oscillations are decreasing when the
temperature increases both for the squeezed light and for
the correlated light.We study the dependence of these
oscillations decreasing on the quadrature squeezing and
correlation coefficient. Also we discuss how the
oscillations of photon distribution function for squeezed
and correlated light deform the Planck distribution formula.
We study the photon
distribution oscillations using the analytical expressions
found in Refs.\cite{Sem},\cite{Dod1}.

\section{Analitical form of photon distribution}
As it was shown in \cite{Sem} the density matrix of the
generic Gaussian state of one mode oscillator corresponding
to the temperature $\beta ^{-1}$ is described by the formula
\beq
\hat \rho =K\hat \rho _{\beta} K\dag =
2\sinh \frac{\beta }{2}\exp [-\beta (
u^{*}a\dag +v^{*}a+\delta ^{*})(ua+va\dag +\delta )+\frac{1}{2}],
\eeq
where
\beq
\hat \rho _{\beta }
=2\sinh \frac{\beta }{2}\exp [-\beta (a\dag a+\frac{1}{2})].
\eeq
For the state (2) we have Planck  distribution formula
\beq
\bar n =\frac{1}{e^{\beta }-1}.
\eeq
The operator $K$ determines the generic linear canonical
transformation of quadratures of the form
\beq
KaK\dag =ua+v{a\dag}+\delta ,
\eeq
where the complex numbers $u$,$v$ and $\delta $ are the
parameters of this transformation satisfying the condition
\beq
|u|^{2}-|v|^{2}=1,
\eeq
which preserves the boson commutation relations of the
operators $a$ and $a\dag $. This transformation, which belongs
to the group $ISp(2,R)$, is determined by these parameters that
label the state (1).
As it was shown in Ref.\cite{Dod1} the five real parameters
$\sigma _{pp},\sigma _{qq},\sigma _{pq},<p>,<q>$ determining
generic Gaussian state of one mode field oscillator are
related to the temperature and canonical transformation parameters
by the relations
\begin {eqnarray}
\sigma _{pp}=E(1+2|v|^{2} +2 Re~ {uv^{*}}),\\
\sigma _{qq}=E(1+2|v|^{2} -2 Re~ {uv^{*}}),\\
\sigma _{pq}=2E Im~ {uv^{*}},\\
<z>=-u^{*}\delta +v\delta ^{*},
\end {eqnarray}
where
\beq
E=\bar n +\frac{1}{2}.
\eeq
The parameters of canonical transformation are expressed in terms
of the squeezing parameter $r$ and correlation parameter
$\theta $ by
\begin {eqnarray}
u=|u|=\cosh r,\\
v=-\sinh r e^{i\theta },
\end {eqnarray}
where
\beq
\theta =\phi _{v} -\pi .
\eeq
In terms of the temperature and of the canonical transformation
parameters the photon distribution function has been
obtained in Ref.\cite{Sem} where it was expressed through
the Hermite polynomials of two variables with the equal
indices.Since these parameters are related to the quadrature
means and dispersion matrix by the formulae written above
the expression of the photon distribution function can be
given in the form obtained
in \cite{Dod1}
\beq
P_{n}=P_{0}\frac{H_{nn}^{\{R\}}(y_{1},y_{2})}{n!},
\eeq
where the probability to have no photon is given by the
formula
\beq
P_0=(d+\frac{1}{2}T+
\frac{1}{4})^{-\frac{1}{2}}\exp [(-<p>^{2}(\sigma _{qq}+
\frac{1}{2})-<q>^{2}(\sigma _{pp}+
\frac{1}{2})+2\sigma _{pq}<p><q>)(\frac {1}{2}+T+2d)^{-1}],
\eeq
the complex parameter $<z>$ is given by the relation
\beq
<z>=2^{-\frac{1}{2}}(<q>+i<p>),
\eeq
and the symmetric matrix $R$ determining Hermite polynomial
has the matrix elements expressed in terms of the quadrature
dispersion matrix as follows
\begin {eqnarray}
R_{11}=(T+2d+
\frac{1}{2})^{-1}(\sigma _{pp}-\sigma _{qq}-2i\sigma _{pq})
=R_{22}^*,\\
R_{12}=(T+2d+\frac{1}{2})^{-1}(\frac{1}{2}-2d).
\end {eqnarray}
Arguments of the Hermite polynomials are of the form
\beq
y_1={y_2}^*=(T-2d-\frac{1}{2})^{-1}[(T-1)<z^*>+
(\sigma _{pp}-\sigma _{qq}+2i\sigma _{pq})<z>].
\eeq
In above formulae the trace $T$ of the quadrature dispersion
matrix and its determinant $d$ are used
\beq
T=\sigma _{pp}+\sigma _{qq}
\eeq
and
\beq
d=\sigma _{pp}\sigma _{qq}-\sigma _{pq}^2.
\eeq
For the pure states the parameter $d$ is equal to 1/4.
The obtained expression for photon distribution function is
convinient for the numerical analysis since there exists the
relation of the Hermite polynomial of two variables in terms
of the usual Hermite polynomial and using this formula the
following photon distribution function for the squeezed and
correlated light subject to thermal noise has been
written down in \cite{Dod1}
\beq
P_{n}=
P_{0}n!\sum_{k=0}^{n}(\frac {R_{11}R_{22}}{4})^{n/2}  (-
\frac {2R_{12}}{\sqrt {R_{11}R_{22}}})^{k}(n-k)!^{-2}k!^{-1}
|H_{n-k}(\frac {R_{11}y_{1}+R_{12}y_{2}}{\sqrt {2R_{11}}})|^{2}.
\eeq
Here the matrix elements of the matrix $R$ are given by the
formulae (17) and (18) and the components $y_{1}$ and $y_{2}$
are given by formula (19). If one calculates the mean value of
photon number $<n>$ corresponding to this distribution the
deformed Planck distribution formula may be obtained \cite{Sem}.
The correction to usual Planck distribution depends on the
squeezing parameter $r$ and the shift parameter $<z>$.This
correction does not depend on the quadrature correlation
coefficient. So we have
\begin{equation}
<n>=\bar n+\sinh ^{2}r(1+2\bar n)+|<z>|^{2}.
\end{equation}
But the width of the photon distribution depends on the
correlation of the quadrature components \cite{Dod2}.
Thus the presence of the oscillations of the photon distribution
function for squeezed and correlated light is related to the
deformation of Planck distribution formula. The correlation
coefficient $k$ of the photon quadratures may be related to the
squeezing parameter $r$ and the phase $\theta $ according
to the formula
\begin{equation}
k=\sin \theta \tanh 2r(1-\tanh ^{2}2r\cos ^{2}\theta )^{-1/2}.
\end{equation}
The modulus of the correlation coefficient is always less than 1.
For the squeezed states the condition of minimisation of
Heisenberg uncertainty relation \cite{Heis} for photon quadratures
is fulfilled
\begin{equation}
\sigma _{pp}\sigma _{qq}=1/4.
\end{equation}
For the correlated states \cite{Kur} the condition of minimisation
of Schr\"odinger uncertainty relation \cite{Schr}
\begin{equation}
\sigma _{pp}\sigma _{qq}-\sigma _{pq}^{2}\ge 1/4,
\end{equation}
is realized,i.e.
\begin{equation}
\sigma _{pp}\sigma _{qq}=\frac {1}{4}\frac {1}{1-k^{2}}.
\end{equation}
For $k=0$ the Schr\"odinger uncertainty relation becomes the
Heisenberg inequality. Thus we have extra physical parameter $k$
 which is the photon quadrature correlation coefficient.

\section{Distribution oscillations versus squeezing,correlation
and temperature}
Using the general formula (22) we will analyse the behaviour of
photon distribution function at different values of the
parameters $r,k,\beta ^{-1}$. For the case of zero quadrature
correlation and zero temperature the photon distribution
function demonstrates the oscillatory behaviour for the large
squeezing.

We present in Fig.1 the plot of photon distribution function
for very low (almost zero) temperature $\beta ^{-1}=10^{-6}$
(solid curve), and for larger temperature
$\beta ^{-1}=5$ (dashed curve), for both cases we have
zero quadrature correlation coefficient $k=0$ and the shift
parameter $<z>= \frac{3+i 2.5}{\sqrt{2}}$~~
(the squeezing parameter $r$ is taken to be $r=5$).

The plot demonstrates the oscillations of photon distribution
function found in Ref.\cite{Sch1}.

If one takes into account
the possible correlation of the photon quadrature components,
assuming the correlation coefficient to be nonzero, the
oscillations of the photon distribution function are available
too. This result has been obtained in Ref.\cite{Kli}

In Fig.2 the plot of photon distribution function is given for
$k=0.9999999835$, $r=5$, $<z>= \frac{3+i 2.5}{\sqrt{2}}$~~ while
$\beta^{-1}=10^{-6}$ (solid), $\beta^{-1}=5$ (dashed).
We see that the oscillations of the photon distribution
function for the squeezed light are sensitive to the presence of
the quadrature correlation since the shape of the curve changes
when the parameter $k$ changes, in fact it oscillates more rapidly
than the previous one.

Fig.3 shows the photon distribution for the case $k=.5066,~
\beta^{-1}=5$,~ $r=5$, $<z>= \frac{3+i 2.5}{\sqrt{2}}$.
One can see the essential difference of the plot from the previous
one. It shows physical relevance of the correlation between photon
quadrature components.

%In Fig.3 we present the plot of the photon distribution function
%with the same squeezing parameter $r=     $,correlation
%coefficient $k=    $ and shift parameter $z=    $, but now the
%temperature is taken to be nonzero, $\beta ^{-1}=       $. We
%see that the amplitude of oscillations becomes less and the photon
%distribution function is more smooth than the curve in Fig.2.

Fig. 4 and 5 present the 3 - dimensional plots of photon
distribution function for the squeezed and correlated light with
the parameters $r=5~ ,\theta=0~ ,z=\frac{3+i 2.5}{\sqrt{2}}$
in Fig.4, and $r=5~ ,\theta=\pi~,z=\frac{3+i 2.5}{\sqrt{2}}$
in Fig.5, as a function of temperature (abscissa) and photon number
(ordinate).
We see that all the amplitudes of the oscillations
of photon distribution for squeezed and correlated light decrease
when the temperature increases. For high temperatures the
oscillations tend to disappear. One could obtain an estimation of
the temperature dependence of the photon distribution function
 (14) for high temperature if the expansion of the density
matrix into power series is used. Then for fixed photon number
$n$ and $\beta \rightarrow 0$ we have approximate expression
\begin{equation}
P_{n}=\beta [1-\beta (n\cosh 2r+\sinh ^{2}r+|\delta |^{2}+1/2)],
\end{equation}
which is valid for not very large $n$. Thus the probability
to have $n$ - photons at high temperatures decreases
proportionally to $\beta $. In the graphs of type presented in
Fig.4 this behaviuor maight be
seen if one takes the plot which may be obtained in the
intersection of the plane crossing the photon number axis in
$n=const$ parallel to the temperature and $P_{n}$ axis. At high
temperatures the decreasing of the photon probability corresponds
just to this law if the temperature is so large that corrections term
in (28) is much less than 1.
Also the correction to this general behaviour
are decreasing with the temperature.

\section{Conclusion}
As we have shown for both kinds of one - mode Gaussian classical
light,namely,for the squeezed light and for the correlated light
the thermal noise supresses the oscillations of photon distribution
function. The presence of oscillations of the photon distribution
function for squeezed and correlated light produces the deformation
of the Planck distribution formula and the correction to the Planck
distribution is dependent on the squeezing parameter.Generalization
of the obtained results for two - mode and polymode squeezed and
correlated light may be done on the base of the expressions for the
photon distribution function of the generic Gaussian light in
terms of multivariable Hermite polynomials found in Ref.\cite{Dod2}.

\section{Acknowledgements}
One of us (V.I.M.)thanks INFN and University of Naples "Federico II"
for the hospitality.

\pagebreak

\centerline{\bf Figure captions}

\vskip 2cm

\noindent Fig.1: Photon distribution for $k=0,~~
<z>= \frac{3+i 2.5}{\sqrt{2}}$
and $\beta^{-1}=10^{-6}$ solid line, $\beta^{-1}=5$ dashed line.

\vskip 2cm

\noindent Fig.2: Photon distribution for $k=0.9999999835,~~
<z>= \frac{3+i 2.5}{\sqrt{2}}$ and $\beta^{-1}=10^{-6}$ solid line,
$\beta^{-1}=5$ dashed line.

\vskip 2cm

\noindent Fig.3: Photon distribution for
$k=0.5066,~~<z>= \frac{3+i 2.5}{\sqrt{2}}$ and $\beta^{-1}=5$.

\vskip 2cm

\noindent Fig.4: 3-dimensional plot of photon distribution function for the
squeezed and correlated light with $r=5,~\theta=0,~
<z>=\frac{3+i 2.5}{\sqrt{2}}$.

\vskip 2cm

\noindent Fig.5: 3-dimensional plot of photon distribution function for the
squeezed and correlated light with
$r=5,~\theta=\pi,~<z>=\frac{3+i 2.5}{\sqrt{2}}$.

\end{document}